\documentclass[10pt,journal=jacsat,manuscript=communication,layout=twocolumn]{achemso}
\usepackage[usenames,dvipsnames]{xcolor}
\usepackage[bookmarks=false,colorlinks]{hyperref}
\hypersetup{
	linkcolor=black, 
	citecolor=black, 
	filecolor=black, 
	urlcolor=black, 
}
\usepackage{amsmath,amssymb}
\usepackage{mdwlist} 
\usepackage{booktabs}
\usepackage{multirow}
\usepackage{amsmath}
\usepackage{lettrine}
\definecolor{col1}{rgb}{0.0, 0.46, 0.8}
\definecolor{col2}{rgb}{0.9, 0.00, 0.30}

\usepackage{lineno}

\usepackage[draft]{changes} 
\usepackage[normalem]{ulem}
\usepackage{tikz}

\newcommand{\mysquare}[1]{\tikz{\filldraw[draw=#1,fill=#1] (0,0)   rectangle (0.7em,0.7em);}}

\definechangesauthor[name={Jiangang}, color=red]{jh}

\title{Pd$_{2}$Se$_{3}$ Monolayer: A Promising Two Dimensional Thermoelectric Material with Ultralow Lattice Thermal Conductivity and High Power Factor}

\author{S. Shahab Naghavi}
\affiliation[Shahid Beheshti University]{Department of Physical and Computational Chemistry, Shahid Beheshti University, G.C., Evin, 1983969411 Tehran, Iran}

\author{Jiangang He}
\affiliation[Northwestern University]{Department of Materials Science and Engineering, Northwestern University,
	Evanston, Illinois 60208, USA} 

\author{Yi Xia}
\affiliation[Argonne National Laboratory]{Center for Nanoscale Materials, Argonne National Laboratory, 9700 South Cass Avenue, Lemont, Illinois 60439, United States} 
\author{C.~Wolverton}

\affiliation[Northwestern University]{Department of Materials Science and Engineering, Northwestern University,
	Evanston, Illinois 60208, USA} 

\begin{document}

\begin{abstract}
A high power factor and low lattice thermal conductivity are two essential ingredients of highly efficient thermoelectric materials. Although monolayers of transition metal dichalcogenides possess high power factors, high lattice thermal conductivities significantly impede their practical applications. Our first-principles calculations show that these two ingredients are well fulfilled in the recently synthesized Pd$_{2}$Se$_{3}$ monolayer, whose crystal structure is composed of [Se$_{2}$]$^{2-}$ dimers, Se$^{2-}$ anions, and Pd$^{2+}$ cations coordinated in a square planar manner. Our detailed analysis of third-order interatomic force constants reveals that the anharmonicity and soft phonon modes associated with [Se$_2$]$^{2-}$ dimers lead to ultra-low lattice thermal conductivities in Pd$_{2}$Se$_{3}$ monolayers (1.5 and 2.9 Wm$^{-1}$K$^{-1}$ along the $a$- and $b$-axes at 300\,K respectively), which are comparable to those of high-performance bulk thermoelectric materials such as PbTe.  Moreover, the ``pudding-mold'' type band structure, caused by Pd$^{2+}$ ($d^{8}$) cations coordinated in a square planar crystal field, leads to high power factors in Pd$_{2}$Se$_{3}$ monolayers. Consequently, both electron and hole doped thermoelectric materials with a considerably high $zT$ can be achieved at moderate carrier concentrations, suggesting that Pd$_{2}$Se$_{3}$ is a promising two-dimensional thermoelectric material.
\\
\\
\textcolor{red}{\textbf{KEYWORDS:}} Pd$_2$Se$_3$ monolayer, thermoelectric, ultralow lattice thermal conductivity, pudding-mold band
\end{abstract}

\maketitle
\lettrine[lines=2]{\color{red}\textbf{T}}{}hermoelectric (TE)
materials  enable an environmentally friendly solution for direct and reversible conversion between heat and electricity. This two-way process has found increasing technological applications, such as  solid-state refrigerators~\cite{rowe1995crc}, flat-panel solar thermoelectric generators~\cite{Kraemer2011}, space power, and recovery of waste heat~\cite{rowe1995crc}.  Nevertheless, for a widespread use of TE materials, their efficiencies need to be significantly improved~\cite{Snyder2008}. The efficiency of TE materials is indexed by the dimensionless figure of merit  $zT\!=\!{S^{2}\sigma\, T}/{(\kappa_{\rm e}+\kappa_{\rm L})}$, where $\sigma$ is electrical conductivity, $S$ is thermopower or Seebeck coefficient, $T$ is absolute temperature, $\kappa_{\rm e}$ and $\kappa_{\rm L}$ are  respectively electrical and lattice thermal conductivities; $S^{2}\sigma$ is usually called power-factor ($\rm PF$).
An effective approach to improve $zT$ is to reduce $\kappa_{\rm L}$, either by searching for materials with intrinsically strong anharmonicity, or by enhancing phonon scattering by phonon engineering~\cite{Toberer2011}, e.g., nanostructuring~\cite{Majumdar777, Xie2012, Poudel2008, Biswas2012}. Likewise, another strategy is to enhance the $\rm PF$ by band structure engineering~\cite{pei2011, Liu2012,Yu2012} or finding a material with a desirable electronic structure, such as small band effective mass and high valley degeneracy~\cite{pei2011}, or the flat-and-dispersive band structure\cite{He2017,Blic2015} (or pudding-mold band).~\cite{Hidetomo2017,Kuroki2007,He2017,Blic2015,isaacs2018inverse} In general, improving the electronic part (i.e., $S$, $\kappa_{\rm e}$, and $\sigma$) is challenging: $S$ and $\sigma$ are generaly inversely related~\cite{Mateeva1998} and $\kappa_{\rm e}$ is proportional to $\sigma$ (Wiedemann-Franz law). Therefore, improving the PF requires the tuning of three conflicting parameters, making the optimization of $zT$ an extremely difficult task.

Alternatively, early theoretical work by Dresselhaus et al.~\cite{Hicks1993, Dresselhaus1999,dresselhaus2007new} and subsequent experimental work~\cite{Harman2002, Venkatasubramanian2001, Hochbaum2008, Zhang2012, Fei2014, Yang2014} suggest that reducing the dimensionality of materials could significantly enhance $zT$. The quantum confinement effect in low dimensional materials significantly increases the density of electronic states, thus increasing PF, and their interfaces/surfaces can effectively scatter heat carrying phonons and thus suppressing $\kappa_{\rm L}$. In fact, a simultaneous increase of $S$ and reduction of $\kappa_{\rm L}$ has been observed in one-dimensional semiconducting materials (Bi$_{2}$Te$_{3}$ nanowires~\cite{Zhang2012}) and many two-dimensional (2D) semiconductors, such as phosphorene monolayers~\cite{Fei2014}, silicene~\cite{Yang2014}, and germanene~\cite{Yang2014}.

In this context, transition-metal chalcogenide monolayers with nonzero band gaps have been intensively studied as promising candidates for 2D TE applications~\cite{Huang2013, Kumar2015, Guo2016PtSe, Wu2014, Yoshida2016, Buscema2013, Li2013, Guo2016, Ding2016, Guo2017, Shafique2017, Jin2016, Sun2015, Wang2017, Wang2015, Zhang2016, Zhang2017}. In particular, transition-metal dichalcogenides (TMDCs) have been the focus of recent studies due to their large $S$~\cite{Wu2014,Yoshida2016,Buscema2013}. A previous study~\cite{Wu2014} found a remarkable enhancement of $S$ in MoS$_{2}$ monolayers (30~mVK$^{-1}$) relative to the bulk phase ($\sim$~7 mVK$^{-1}$). Despite improving $S$, a sizable $zT$ has not been yet realized in TMDCs owing to their high $\kappa_{\rm L}$~\cite{Huang2013,Kumar2015,Guo2016PtSe,Peng2015} rooted in the covalent nature of bond between transition metal and chalcogenide atoms.\cite{Chhowalla2013,Yun2012} The calculated $\kappa_{\rm L}$ of  MoS$_{2}$, MoSe$_{2}$ and WSe$_{2}$, based on density functional theory (DFT), at 300 K are respectively 140, 80 and 40 $\rm  Wm^{-1}K^{-1}$~\cite{Kumar2015}, which are in good agreement with the measured $\kappa_{\rm L}$.


\section{Structure}
\label{sec:structure}

\begin{figure}[htp!]
	\centering
	\includegraphics[width=1.0\linewidth]{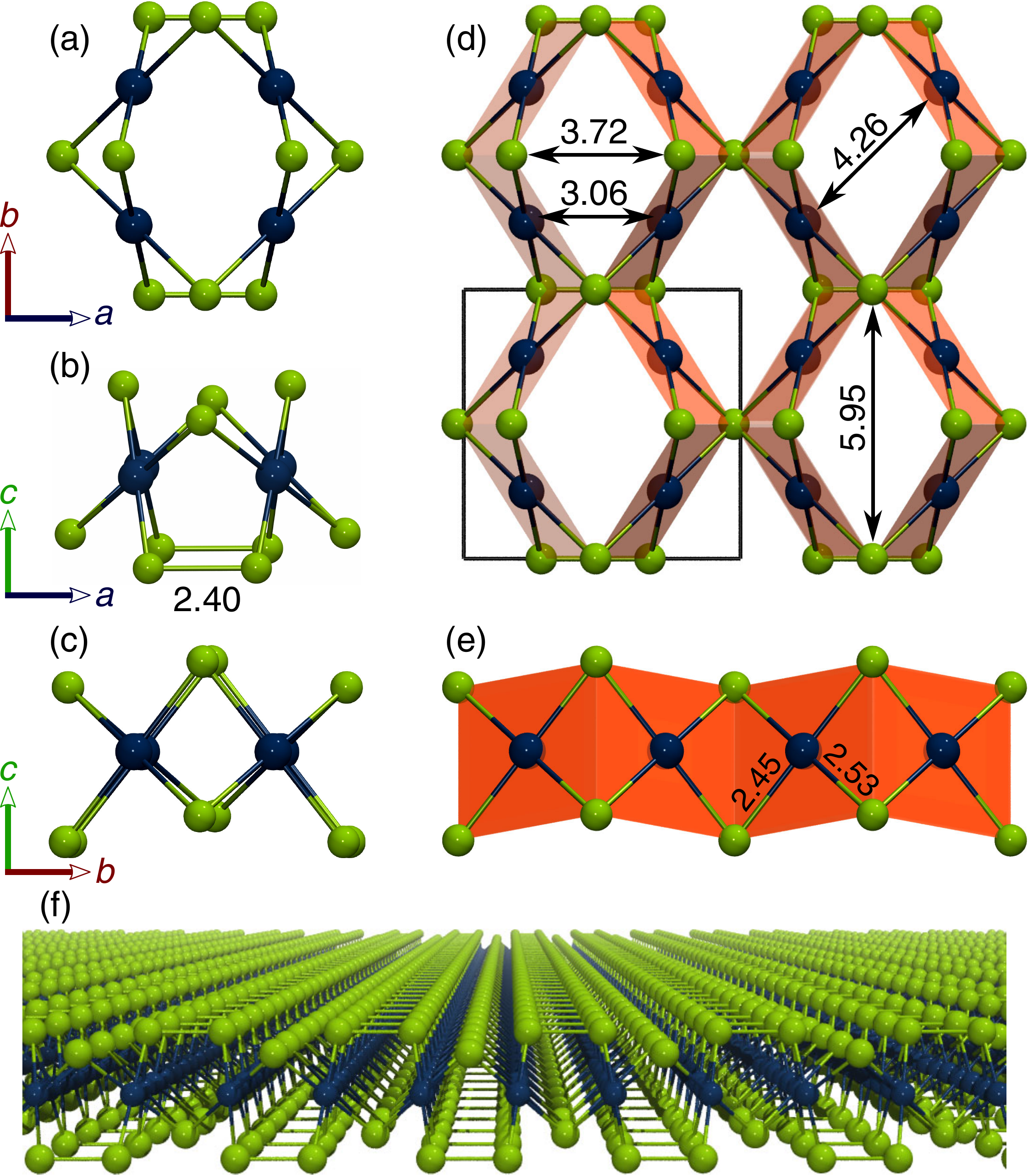}
	\caption{(a)--(f) Different views of monolayer Pd$_{2}$Se$_{3}$ structure. The square planar units are highlighted in orange.}
\label{FIG:1}
\end{figure}

Transition-metal \textit{tri-}chalcogenide (TMTCs) monolayers~\cite{Island2014, Lipatov2015, Pawbake2015} usually offer more complex atomic configurations and therefore more tortuous phonon paths~\cite{Zhang2017} by incorporating both $X_{2}^{2-}$ dimers and $X^{2-}$ ($X$=S and Se) anions in their crystal structures. For instance, the $\kappa_{\rm L}$ of TiS$_{3}$~\cite{Zhang2017} at 300\,K ($\sim$\,10\,$\rm Wm^{-1}K^{-1}$) is much lower than that of WSe$_{2}$ (40 $\rm Wm^{-1}K^{-1}$), even though both Ti and S atoms have much smaller atomic masses than W and Se. Nevertheless, such complex crystal structures are scarce in monolayers.
Recently, Lin {\it{et.\,al.}}~\cite{Lin2017} successfully synthesized a novel semiconducting Pd$_{2}$Se$_{3}$ monolayer with a unique crystal structure. The Pd$_{2}$Se$_{3}$ monolayer was synthesized by the fusion of two monolayers of PdS$_{2}$, though no Pd$_{2}$Se$_{3}$ bulk compound is yet reported. Due to the complex crystal structure ([Se$_{2}$]$^{2-}$ and Se$^{2-}$) and a large void in the monolayer, which is also not common among 2D materials, Pd$_{2}$Se$_{3}$ monolayers are expected to have  a low $\kappa_{\rm L}$. Moreover, the presence of pudding-mold type band structure~\cite{Hidetomo2017,Kuroki2007,He2017,Blic2015} could lead to a high PF in the Pd$_{2}$Se$_{3}$ monolayer.

In this work, we use first-principles DFT band structure, anharmonic phonon calculations, and Boltzmann transport theory\cite{allen1996boltzmann}, to provide a comprehensive study on the electronic and phonon transport properties of the Pd$_{2}$Se$_{3}$ monolayer. Our results show that Pd$_{2}$Se$_{3}$ monolayers have much lower $\kappa_{\rm L}$ and higher PF than all the previously reported transition-metal dichalcogenides~\cite{Kumar2015} and trichalcogenides~\cite{Zhang2017}, and thus possesses an overall better TE performance.

All the DFT calculations were performed using the projector-augmented wave (PAW) method~\cite{PAW1,PAW2} as implemented in the Vienna Ab-initio Simulation Package (\texttt{VASP})~\cite{VASP1,VASP2}. A plane wave basis set with energy cutoff of 350~\,eV and the generalized gradient approximation of Perdew-Burke-Ernzerhof (PBE)\cite{PBE} to the exchange-correlation functional were used. A $12\times12\times1$ $k$-mesh is used to sample the first Brillouin zone. All structures were fully relaxed with respect to lattice vectors and atomic positions until the forces on each atom are less than 0.1~meV {\AA}$^{-1}$. We found that the spin-orbit coupling (SOC) does not alter the dispersion of energy levels close to the Fermi level (see Figure\,S3) and therefore SOC was not included in our calculations. Electrical transport properties, i.e., $S$, $\sigma$ and $\kappa_{\rm e}$, were calculated using the Boltzmann transport theory within the constant relaxation time approximation as implemented in \texttt{BoltzTrap}.\cite{BoltzTrap} The reciprocal space was sampled with a dense $k$-grid of  $38\times38\times1$, to enable accurate Fourier interpolation of the Kohn-Sham eigenvalues. The $\kappa_{\rm e}$ was  calculated using the Wiedemann-Franz law ($\kappa_{\rm e}$ = $L\sigma{T}$) with $L = 2.45\times \rm 10^{-8}\,W\Omega K^{-2}$. Using the \texttt{ShengBTE} code~\cite{ShengBTE_1,ShengBTE_2,ShengBTE_3}, the $\kappa_{\rm L}$ is computed by solving the Boltzmann transport equation of phonons with the second- and third-order interatomic force constants (IFCs) as input. The second-order IFCs were calculated by the \texttt{Phonopy} code~\cite{Phonopy} using a $6\times6\times1$ supercell with $2\times2\times1$ $k$-point sampling. A $3\times 3\times 1$ supercell with $4\times 4\times 1$ $k$-point sampling was used to obtain third-order IFCs uing \texttt{ShengBTE} code~\cite{ShengBTE_1,ShengBTE_2,ShengBTE_3}. A well-converged $q$-mesh ($30\times30\times1$) was used to calculate $\kappa_{\rm L}$ and related phonon properties.

Experimentally, Pd$_{2}$Se$_{3}$ monolayers have been synthesized by interlayer fusion of two defective PdSe$_{2}$ layers.~\cite{Lin2017} The monolayer is stable when exposed to air and at elevated temperatures~\cite{Lin2017}. Its crystal structure has an inversion center with the point group of $D_{2h}$ ($mmm$) and DFT calculated lattice parameters are 6.12 and 5.95\,{\AA}. As shown in Figure~\ref{FIG:1}, the Pd atom has a square-planar coordination formed by [Se$_2$]$^{2-}$ and Se$^{2-}$. Such a coordination geometry is common among transition-metal complexes with the $d^{8}$ electronic configuration~\cite{He2017,isaacs2018inverse}, suggesting that the oxidation state of  Pd is  2+ in Pd$_{2}$Se$_{3}$. Therefore, the coexistence of [Se$_2$]$^{2-}$ dimers and Se$^{2-}$ anions leads to oxidation states of the stoichiometric compound as [Pd$^{2+}$]$_2$[Se]$^{2-}$[Se$_{2}$]$^{2-}$.  The formation of [Se$_{2}$]$^{2-}$  dimers are supported by DFT calculated electron localization function (ELF) shown in Figure~\ref{FIG:4} (e), where the attractors (red area) on the midpoint of two selenide atoms indicate the covalent nature of bonding.  The calculated Se--Se bond length of 2.4 {\AA} (see Figure\,\ref{FIG:1}) is consistent with a fully oxidized two-center two-electron [Se$_2$]$^{2-}$ dimer~\cite{OKeeffe1992}. Therefore, there are two types of Pd--Se bonds in the Pd$_2$Se$_3$ monolayer: the longer Pd--Se bond formed between Pd$^{2+}$ cation and Se$^{2-}$ anion (2.53~\AA) and the shorter one formed between Pd$^{2+}$ and ${\rm [Se_{2}]^{2-}}$ dimer (2.45\AA). The resulting crystal structure is thus complex with large rhombus-shape voids, as shown in Figure~\ref{FIG:1}.

Slack's theory~\cite{Slack1973} reveals that four factors lead to low $\kappa_{\rm L}$: i) complex crystal structure, ii) high average mass, iii) weak interatomic bonding, and iv) anharmonicity. As will be discussed later, the complex crystal structure of Pd$_{2}$Se$_{3}$  features all those key characteristics, in particular, a strong anharmonicity stemming from the [Se$_2$]$^{2-}$ dimer. 


\begin{figure}[htb!]
    \centering
    \includegraphics[width=1.0\linewidth]{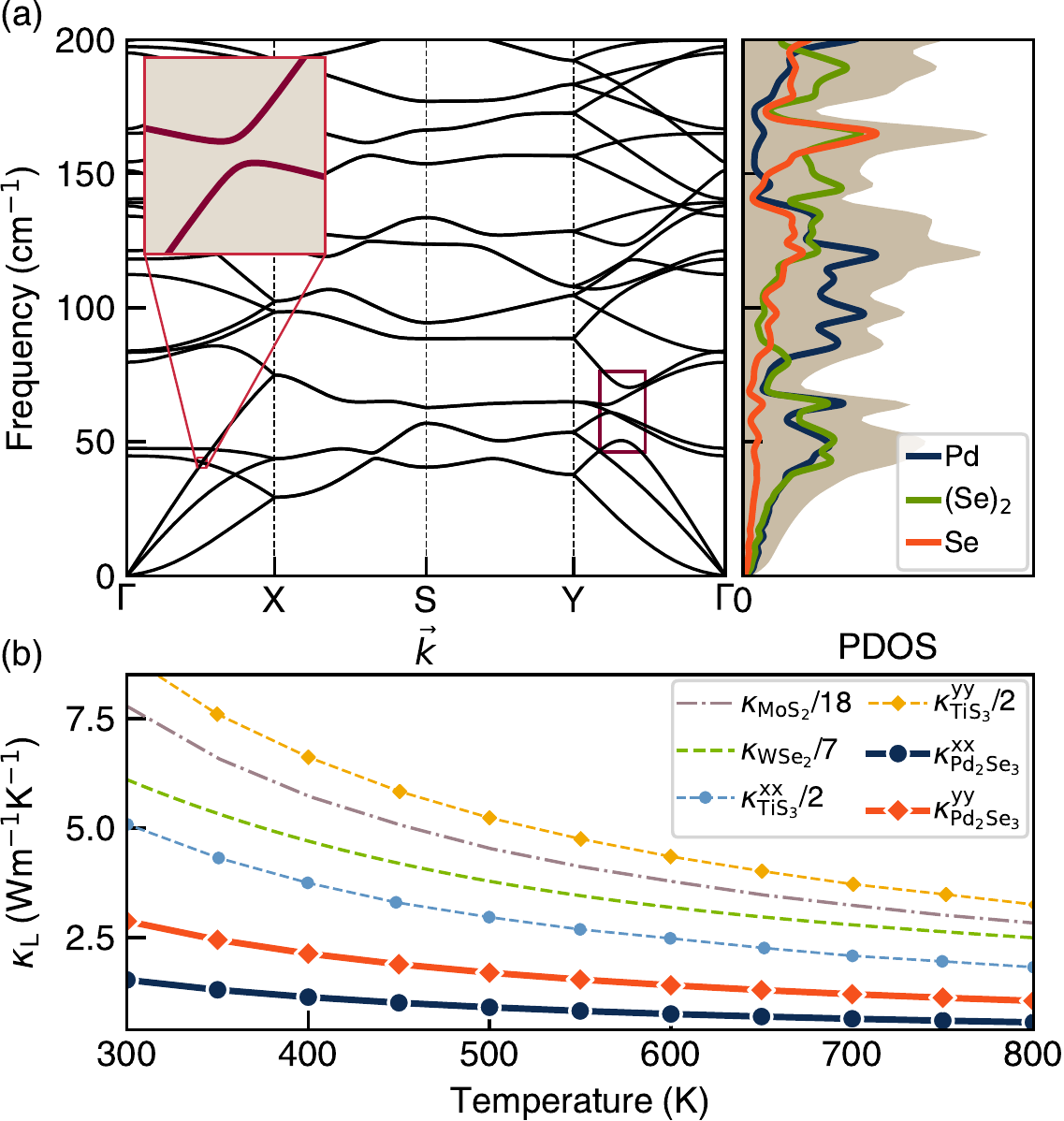}
    \caption{(a) Phonon dispersion and the density of states of Pd$_{2}$Se$_{3}$ with the highlighted avoided crossing bands. The density of states indicates that low-frequency modes are mainly composed of Se$_{2}$ dimers and heavy Pd atoms. (b) Calculated $\kappa_{\rm L}$ of Pd$_{2}$Se$_{3}$ compared to MoS$_{2}$, WSe$_{2}$ and TiS$_{3}$ (notice that their $\kappa_{\rm L}$ is divided to an arbitrary number to fit into the figure window). The $\kappa_{\rm L}$ of Pd$_{2}$Se$_{3}$ is about 90, 27, 6 times lower than  MoS$_{2}$~\cite{Gandi2016}, WSe$_{2}$\cite{Kumar2015}, and TiS$_{3}$~\cite{Zhang2017} respectively. The calculated $\kappa_{L}$ of MoSe$_{2}$\cite{Kumar2015}, which is not shown here, at T= 300\,K is about 80\,Wm$^{-1}$K$^{-1}$. The atom projected $\kappa_{\rm L}$ (see Figure\,S9 in Supplementary Materials) indicates that Se$_{2}$ dimers and Pd atoms carry the most heat.}
\label{FIG:2}
\end{figure}

Figure~\ref{FIG:2}\,(a) displays the phonon dispersion of the Pd$_{2}$Se$_{3}$ monolayer. The frequencies of all modes are positive in the whole Brillouin zone (BZ), implying that the Pd$_{2}$Se$_{3}$ monolayers are dynamically stable.
\begin{figure*}[htb!]
    \centering
    \includegraphics[width=1.0\linewidth]{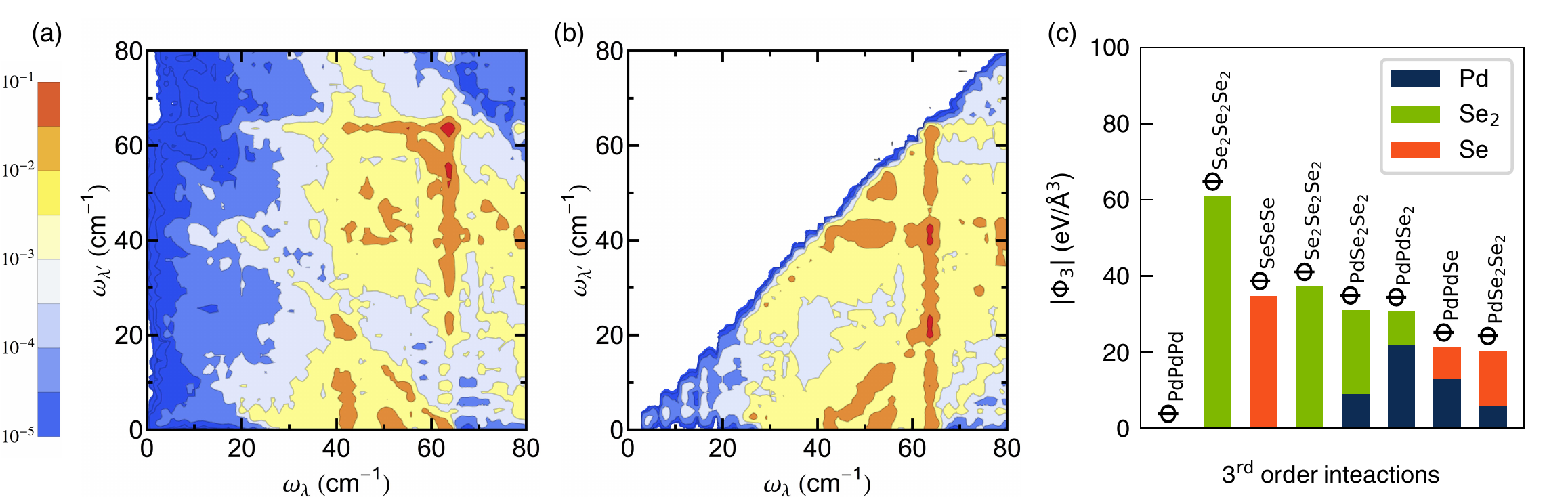}
    \caption{Contour plot of phonon scattering rate of the first phonon mode $\lambda$ induced by the second phonon mode $\lambda^{\prime}$ in the three phonon processes, namely, absorption (a) and emission (b) processes in which crystal momentum and energy are conserved. (c) Norm of the calculated third-order interatomic force constants, indicating the magnitude of anharmonicity.}
\label{FIG:3}
\end{figure*} 
The zone-boundary frequencies along $\Gamma$--$X$ and $\Gamma$-$Y$ are as low as 30 and 40\,cm$^{-1}$. As seen in Figure~\ref{FIG:2}, the acoustic bands, which play a dominant role in lattice heat transfer, have frequencies from 0 to 70\,cm$^{-1}$, and are mainly localized on [Se$_2$]$^{2-}$ dimers and Pd$^{2+}$ cations. Although Se$^{2-}$ is light and its associated vibrations appear at high-frequency regions, the [Se$_2$]$^{2-}$ dimer acts like a heavy atom participating in low-frequency vibration modes.

The $\kappa_{\rm L}$ for each direction (i.e., $a$- and $b$-axes) is proportional to the square of the phonon group velocity along the respective direction\cite{mingo2014ab}. The phonon group velocities of Pd$_{2}$Se$_{3}$ for the out-of-plane acoustic (ZA), transverse acoustic (TA), and longitudinal acoustic (LA) modes in the long-wavelength limit are listed in Table~\ref{TAB:VEL}. The sound velocities of Pd$_{2}$Se$_{3}$ for all the acoustic branches are lower than those of MoS$_{2}$ and TiS$_{3}$~\cite{Zhang2017}, suggesting a lower $\kappa_{\rm L}$ in the Pd$_{2}$Se$_{3}$ monolayer. On the other hand, the avoided crossing between the optical and acoustic modes is clearly seen in Figure~\ref{FIG:2}\,(a) along $\Gamma$--X and $\Gamma$--Y directions. The sizable gap at the avoided crossing point indicates a high coupling strength (hybridization) between optical and acoustic modes, which significantly increases the phonon scattering rates and reduces acoustic mode velocities, and thus leads to the low $\kappa_{\rm L}$.

For the quantitative description of $\kappa_{\rm L}$, we use first-principles calculations in conjunction with the self-consistent iterative solution of the Boltzmann transport equation (BTE) for phonons as implemented in \texttt{ShengBTE}~\cite{ShengBTE_1}. The calculated $\kappa_{\rm L}$ as a function of temperature along the $a$- and $b-$directions are shown in Figure~\ref{FIG:2}\,(b). The lattice thermal conductivity along the $a$ ($\kappa_{\rm L}^{\rm xx}$) and $b$-axes ($\kappa_{\rm L}^{\rm yy}$) are 1.5 and 2.85 Wm$^{-1}$K$^{-1}$ at 300\,K, respectively. They are comparable to those of high $zT$ bulk materials such as PbTe, but much lower than other TMCs monolayers such as MoS$_{2}$ (140 Wm$^{-1}$K$^{-1}$), TiSe$_{3}$ (10 Wm$^{-1}$K$^{-1}$), and WSe$_{2}$ (42 Wm$^{-1}$K$^{-1}$). The calculated cumulative $\kappa_{\rm L}$  with respect to mean free path is shown in Figure~S7. The result indicates that the $\kappa_{\rm L}$ can be further reduced by decreasing grain size of the polycrystal; for instance, at the size of 75\,nm the $\kappa_{\rm L}^{\rm xx}$ and $\kappa_{\rm L}^{\rm yy}$ of the Pd$_{2}$Se$_{3}$ monolayer drop by 50\%.  

\begin{table}[htb!]
\caption{Calculated group velocities of ZA, TA and LA phonons near the $\Gamma$ point for the Pd$_{2}$Se$_{3}$ monolayer along the $a-$ and $b-$direction compared with TiS$_{3}$ and MoS$_{2}$.}
\centering
\resizebox{\columnwidth}{!}{\begin{tabular}{l@{\hskip 0.150in}c@{\hskip 0.150in}c@{\hskip 0.150in}c@{\hskip 0.150in}c@{\hskip 0.0in}}
                              \toprule
                              
Monolayer                         & Direction & ZA (Km/s) & TA(Km/s) & LA(km/s) \\
\midrule
\multirow{2}{*}{Pd$_{2}$Se$_{3}$} & $x$       & 0.51      & 2.25     & 3.12     \\
                                  & $y$       & 0.50      & 2.27     & 3.19     \\
\multirow{2}{*}{TiS$_{3}$}        & $x$       & 0.88      & 3.01     & 5.43     \\
                                  & $y$       & 1.11      & 2.31     & 6.16     \\
MoS$_{2}$                         & $x$, $y$  & 1.40      & 3.96     & 6.47     \\
\bottomrule
\end{tabular}}
\label{TAB:VEL}
\end{table}

A comparison of $\kappa_{\rm L}$ for two selenide based TMCs monolayers, Pd$_{2}$Se$_{3}$ and  WSe$_{2}$, is illuminating. The atomic mass of the Pd (106.42) is much smaller than the W (183.4) and thus one might expect a higher $\kappa_{\rm L}$ for the Pd$_{2}$Se$_{3}$. Nevertheless, as seen in Figure~\ref{FIG:2} the calculated $\kappa_{\rm L}$ of the Pd$_{2}$Se$_{3}$ monolayer is more than 20 times lower than WSe$_{2}$. We next explore the origin of this behavior.

From the cumulative $\kappa_{\rm L}$ of phonon frequency (see Figure~S8 in Supplementary Materials) we see that lattice heat transport is dominated by phonon modes with frequencies less than 80\,cm$^{-1}$. In Figure~\ref{FIG:3} (a) and (b) we show the scattering rates associated with these low-lying phonon modes from three-phonon interactions, namely, the absorption ($\Gamma^{+}$: $\lambda+\lambda^{\prime}\to \lambda^{\prime \prime}$) and emission ($\Gamma^{-}$:$\lambda \to \lambda^{\prime} + \lambda^{\prime \prime}$) processes. Different colors in the scattering rates plot show the scattering magnitude of the first phonon mode ($\lambda$) induced by the second phonon mode ($\lambda^{\prime}$). In the absorption process, a low-frequency phonon mode contributes to other low-frequency phonon modes, giving rise to a high-frequency optical mode. In the emission process, the phonon mode is only allowed to decompose into a lower-frequency mode, thus restricting the second phonon mode ($\lambda^{\prime}$) in the right lower triangle. Both processes satisfy energy and crystal-momentum conservation. 
Figure~\ref{FIG:3} (a) shows the strong scattering of acoustic modes through combination with low-lying optical modes ($\omega$ $\approx$ 40\,cm$^{-1}$), which is near the avoided crossing, consisting with high scattering rates due to avoided crossing bands. Phonon modes with frequencies ranging from 40 to 70\,cm$^{-2}$, as shown in Figure~\ref{FIG:3} (a) and (b), are heavily scattered in both absorption and emission processes, indicating that the presence of low-lying optical modes significantly enhances overall phonon scattering rates.

To specify the role of each atom in the observed low $\kappa_{\rm L}$, we calculate the atom projected $\kappa_{\rm L}$, as shown in Figure~S9 in Supplementary Materials. [Se$_2$]$^{2-}$ dimers and Pd$^{2+}$ cation are largely responsible (90\% in total) for heat transport in the Pd$_{2}$Se$_{3}$ monolayer, whereas the contribution of Se$^{2-}$ is negligible. We also calculate the norm of third-order IFCs defined as $\Phi_{mnl}= \frac{\partial^{3} E}{\partial u_{m}\partial u_{n}\partial u_{l}}$ ($E$ and $u$ are the total energy and atom displacement for different atom species $m$, $n$, and $l$). Since the phonon scattering rates are roughly proportional to $|\Phi|^2$\cite{ShengBTE_1,Mahan1996}, a high absolute value of $\Phi_{mnl}$ suggests a large anharmonicity. As shown in Figure~\ref{FIG:3}\,(c), $\Phi_{\rm Se_{2}Se_{2}Se_{2}}$ indicates a large anharmonicity associated with [Se$_2$]$^{2-}$ dimers, which is much higher than Pd$^{2+}$ and Se$^{2-}$ anions. Therefore, the low $\kappa_{\rm L}$ is a combined effect of strong anharmonic phonon-phonon interactions and small group velocities, stemming from the formation of [Se$_2$]$^{2-}$ dimers. This answers the question on why Pd$_{2}$Se$_{3}$ has a much lower $\kappa_{\rm L}$ than other TMCs, where such dimers do not form.

\section{Electronic}
\label{sec:electronic}

\begin{figure*}[htb!]
	\centering
	\includegraphics[width=1.0\linewidth]{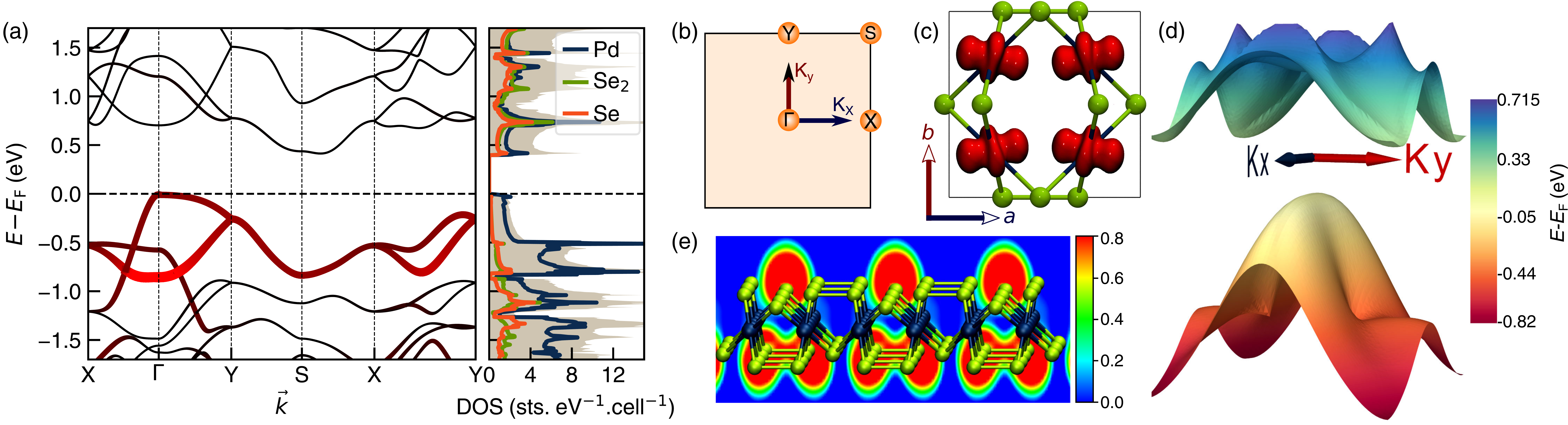}
	\caption{(a) DFT (PBE) orbital projected band structure (line width and red color correspond to the contribution of Pd-4$d_{z^{2}}$ orbitals) and projected density of states (PDOS). (b) Schematic Brillouin zone, (c) decomposed charge desnity of the top of the valence band, (e) electron localization function (ELF) viewed along $b$-axis, and (d) three dimensional band structure of Pd$_{2}$Se$_{3}$.}
\label{FIG:4}
\end{figure*}

As already mentioned, in the Pd$_{2}$Se$_{3}$ monolayers the Pd$^{2+}$ cation has a $d^{8}$ electronic configuration and a square planar crystal field, under which $d$ obritals split into four energy levels, $d_{xz}$/$d_{yz}$, $d_{z^{2}}$, $d_{xy}$, and $d_{x^{2}+y^{2}}$ from low to high energy. On the other hand, the packing of square planar units in the crystal lattice induces a weak interaction between the nearby Pd$^{2+}$ cations, separated by 3.06\,{\AA}, via $d_{z^{2}}$ obritals as seen in Figure~\ref{FIG:4} (c). This interaction switches the energy levels of $d_{z^{2}}$ with $d_{xy}$ (see Figure~S1 in Supplementary Materials). Note that in the Pd$_{2}$Se$_{3}$ monolayers, the $d_{z^{2}}$ orbital almost lies along the $a$-axis. Due to the strong crystal field splitting associated with square planar geometry, the low spin state is always preferred in Pd$^{2+}$. As a result, the four low-energy levels are occupied by the eight electrons of Pd$^{2+}$ ($d^{8}$) cations, where the $d_{z^{2}}$ is the highest occupied orbitals (the top of the valence band). Therefore the Pd$_{2}$Se$_{3}$ monolayer is a band insulator, in which the top of the valence band is mainly composed of $d_{z^{2}}$ orbital, as seen in Figure~\ref{FIG:4}\,(a), and the bottom of the conduction band is largely from $d_{x^{2}+y^{2}}$ (see Figure~S2).

As the overlap of $d_{z^{2}}$ orbital with the $p_{x}$ and $p_{y}$ orbitals of the nearest anion is negligible, a relatively flat band  along the $b$-axis is expected. As seen in Figure~\ref{FIG:4}\,(a) and (b), the valence band maximum along the $\Gamma-Y$ is relatively flat, affording a high density of states (DOS) near the Fermi level. On the other hand, a proper overlap between Pd$^{2+}$ $d_{z^{2}}$  orbitals along the $a$-axis leads to a very dispersive band along the $X-\Gamma$ direction, indicating a small band effective mass and therefore high carrier mobility. This type of band structure, known as the flat-and-dispersive or ``pudding-mold'' type band structure, has been found in high-performance bulk TE materials\cite{isaacs2018inverse} such as Na$_{x}$CoO$_{2}$~\cite{Hidetomo2017, Kuroki2007}, Bi$_2$PdO$_4$~\cite{He2017}, and some full Heusler compounds.\cite{Blic2015,he2016ultralow} This actually coincides with the idea proposed by Mahan and Sofo~\cite{Mahan1996} that ``\emph{we have to search for materials where the distribution of energy carriers is as narrow as possible, but with high carrier velocity in the direction of the applied electric field.}''. A highly dispersive band (small band effective mass) around the Fermi level gives rise to a large $\sigma$, while a sharp increase in the density of states owing to the presence of a flat band (large band effective mass) usually leads to a large $S$~\cite{Mahan1996}. In  the case of the Pd$_{2}$Se$_{3}$ monolayers, the calculated effective masses ($m^{*}$) for holes  along the dispersive band is about 0.17\,m$_{0}$  and for the flat band is 9.14\,m$_{0}$, thus a high $S$ and $\sigma$ are expected along the $a$-axis.                                
                                                                               
\begin{figure}[htb!] \centering                                                                                                                      
\includegraphics[width=1.0\linewidth]{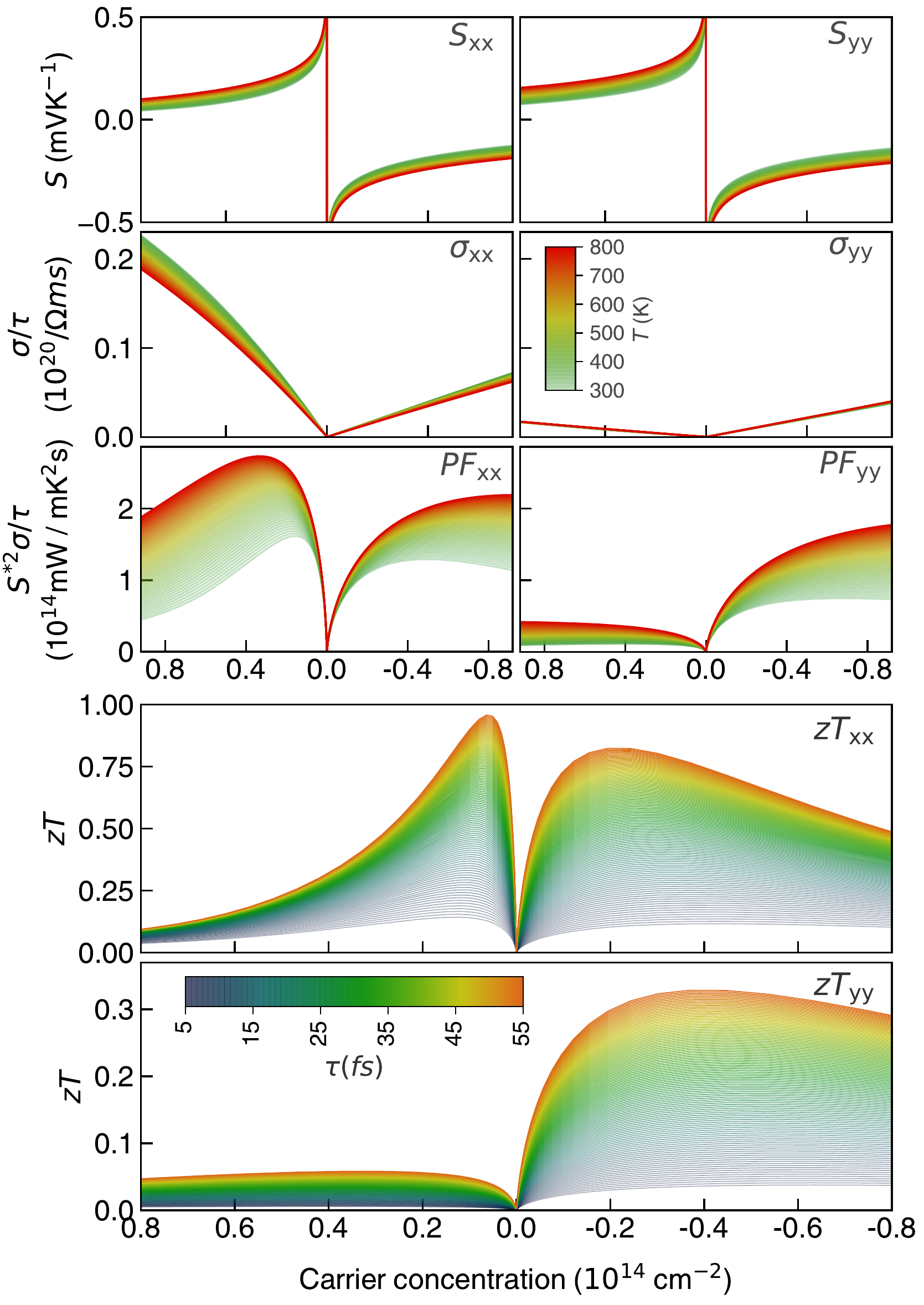}                                                                             
        \caption{The calculated transport coefficients (PF, $S$, $\sigma$)  and $zT$ of Pd$_{2}$Se$_{3}$ as a function of carrier concentrations with respectively various temperatures and electron relaxation time ($\tau$) at 300 K.}
\label{FIG:5}
\end{figure}

In Figure~\ref{FIG:5}, we plot the calculated electronic transport coefficients for hole ($p$-type) and electron ($n$-type) doped 
Pd$_{2}$Se$_{3}$ monolayers at varying temperatures. The calculated PF along the $a$-axis for both $p$-type and $n$-type systems,  assuming $\tau=1\times 10^{-14}$\,s and T=300\,K, are respectively 1.61 and 1.29 mW/mK$^{2}$. Along the $b$-axis, while the PF of $n$-type is large (0.7 mW/mK$^{2}$), the $p$-type is quite small (0.1 mW/mK$^{2}$). Thus, a large PF anisotropy is established with a dominant PF along the $a$-axis in the $p$-type Pd$_{2}$Se$_{3}$ monolayers. Using the same electronic relaxation time, the maximum PF for MoS$_{2}$, MoSe$_{2}$, WSe$_{2}$, and TiS$_{3}$ at the same temperature are $\approx$ 1.8 mW/mK$^{2}$ ($n$-type)\cite{Babaei2014}, 0.8 mW/mK$^{2}$ ($n$-type)\cite{Kumar2015}, 1.7 mW/mK$^{2}$ ($n$-type)\cite{Kumar2015} and 1.8 mW/mK$^{2}$ ($n$-type)~\cite{Zhang2017}, respectively. The PF of Pd$_{2}$Se$_{3}$ monolayers, 1.61  mW/mK$^{2}$ ($p$-type) and 1.21  mW/mK$^{2}$ ($n$-type), is comparable to these TMCs, while its $\kappa_{\rm L}$ is one to two orders of magnitudes lower (see Figure~\ref{FIG:2}). Therefore, a larger $zT$ for the Pd$_{2}$Se$_{3}$ monolayers is expected.


In Figure~\ref{FIG:5} we calculate the $zT$ at the varying relaxation times  within a reasonable            
range~\cite{Gonzalez-Romero2017} from 5 to $55\times 10^{-15}$\,s. The $zT$ of $p$-type Pd$_2$Se$_3$ monolayers along the $a$-axis is about 16\% larger than the $n$-type one. Depending on the relaxation time the calculated $n$-type $zT$ values vary between 0.15 to $\approx$ 1. Along $b$-axis, a large difference between $n$- and $p$-type $zT$ values is found.
Using the same $\tau$, our calculated $zT$ values are much larger than the previously reported TMCs such as TiS$_{3}$~\cite{Zhang2017}, MoSe$_{2}$, WSe$_{2}$~\cite{Kumar2015}, PtSe$_{2}$~\cite{Guo2016PtSe}. Our calculations suggest that Pd$_{2}$Se$_{3}$ monolayers are promising TE material in both $n$-type and  $p$-type applications.


In conclusion, we investigated the electronic structure, phonon, and electron and phonon transport properties of the recently synthesized Pd$_{2}$Se$_{3}$ monolayers by the means of first-principles calculations and Boltzmann transport theory. Our results demonstrate that the Pd$_{2}$Se$_{3}$ monolayers possess a much lower lattice thermal conductivity than other TMC monolayers, e.g., MoS$_{2}$, MoSe$_{2}$, WSe$_{2}$, Ti$_{2}$Se$_{3}$. Detailed analysis of third-order force constants indicates that the anharmonicity and soft phonon modes associated with [Se$_2$]$^{2-}$ dimers are responsible for the low lattice thermal conductivity of Pd$_{2}$Se$_{3}$. On the other hand, the ``pudding-mold'' type band structure, originating from the square-planar coordinated Pd$^{2+}$ cation, offers a high power factor. An extremely low lattice thermal conductivity in conjunction with a high power factor leads, of course, to the superior TE performance in the Pd$_{2}$Se$_{3}$ monolayer. Our results suggest the Pd$_{2}$Se$_{3}$ monolayer is a promising two-dimensional thermoelectric material with a high $zT$ for both hole and electron doping.

\vspace{1cm}
\noindent \mysquare{red} \textcolor{red}{\textbf{ACKNOWLEDGMENTS}} \\
J. H. and C. W. (electronic structure, thermoelectric calculations and analysis) acknowledge support by the U.S. Department of Energy, Office of Science and Office of Basic Energy Sciences, under Award No. DE-SC0014520. Y.X (analysis of phonon calculations) acknowledges the Center for Nanoscale Materials, an Office of Science user facility, supported by the U. S. Department of Energy, Office of Science, Office of Basic Energy Sciences, under Contract No. DE-AC02-06CH11357. The authors acknowledge computing resources provided on Blues, a high-performance computing cluster operated by the Laboratory Computing Resource Center at Argonne National Laboratory.

\providecommand{\latin}[1]{#1}
\makeatletter
\providecommand{\doi}
  {\begingroup\let\do\@makeother\dospecials
  \catcode`\{=1 \catcode`\}=2 \doi@aux}
\providecommand{\doi@aux}[1]{\endgroup\texttt{#1}}
\makeatother
\providecommand*\mcitethebibliography{\thebibliography}
\csname @ifundefined\endcsname{endmcitethebibliography}
  {\let\endmcitethebibliography\endthebibliography}{}

\end{document}